\documentclass[aps,pre,showpacs,twocolumn,superscriptaddress,letterpaper]{revtex4}
\usepackage{mathrsfs}
\usepackage{amsmath}
\usepackage{bm}
\usepackage{graphicx}
\usepackage{hyperref}

\begin{document}


\title{Unconventional Ballooning Structures for Toroidal Drift Waves}
\author{Hua-sheng Xie}
\email[]{Email: huashengxie@gmail.com} \affiliation{Institute for
Fusion Theory and Simulation, Department of Physics, Zhejiang
University, Hangzhou, 310027, People's Republic of China}
\author{Yong Xiao}
\email[]{Email (Corresponding author): yxiao@zju.edu.cn}
\affiliation{Institute for Fusion Theory and Simulation, Department
of Physics, Zhejiang University, Hangzhou, 310027, People's Republic
of China}
\date{\today}

\begin{abstract}
With strong gradients in the pedestal of high confinement mode
(H-mode) fusion plasmas, gyrokinetic simulations are carried out for
the trapped electron and ion temperature gradient modes. A broad
class of unconventional mode structures is found to localize at
arbitrary poloidal positions or with multiple peaks. It is found
that these unconventional ballooning structures are associated with
different eigen states for the most unstable mode. At weak gradient
(low confinement mode or L-mode), the most unstable mode is usually
in the ground eigen state, which corresponds to a conventional
ballooning mode structure peaking in the outboard mid-plane of
tokamaks. However, at strong gradient (H-mode), the most unstable
mode is usually not the ground eigen state and the ballooning mode
structure becomes unconventional. This result implies that the
pedestal of H-mode could have better confinement than L-mode.
\end{abstract}

\pacs{52.35.Py, 52.30.Gz, 52.35.Kt}

\maketitle

Although numerous theoretical models have been
suggested\cite{Wagner2007}, a yet unexplained phenomenon in tokamak
fusion plasmas, is the transition of low (L) to high (H) confinement
states, where H-mode\cite{Wagner1982} has significant better
confinement property than that of the L-mode. Understanding of the
H-mode physics is not only important to make controlled fusion more
feasible, but also that the existence of and transitions among
multi-equilibrium states are important fields of nonlinear physics
in laboratory and the Universe. Drift wave turbulence is one of the
major causes that leads to the anomalous transport widely observed
in fusion and space plasmas\cite{Horton1999,Hasegawa1969}. In order
to control the turbulent transport, it is crucial to understand the
underlying transport mechanism, which may vary for different types
of instability that drive the turbulence. The correlation time and
length are found to be closely related to the mode structure of the
turbulence\cite{Lin2007}. Therefore, the mode structure of the
turbulence has a significant effect on the transport
level\cite{Xiao2010}.

In this Letter, we show that the linear properties of two major
types of electrostatic micro-instabilities\cite{Horton1999}, namely
the trapped electron mode (TEM) and ion temperature gradient (ITG)
mode, are completely different in the H-mode (strong gradient) and
L-mode (weak gradient) stages. With the conventional weak gradient,
the mode structures for drift wave instabilities such as the ITG and
TEM are of ballooning type, peaking at the outboard mid-plane of the
tokamak (c.f., \cite{Xie2012,Dickinson2014}). This type of solution
has been intensively studied using the
ballooning-representation\cite{Connor1978,Connor1987} by reducing
one two-dimensional (2D) real space eigen mode equation
for the drift waves to two one-dimensional (1D) ballooning
space eigen mode equations. For the 2D case we solve the
eigen equation in the poloidal plane. For the 1D case we solve the
eigen equation in the parallel direction. The most unstable
solutions in the ballooning space found in the past have usually the
ballooning-angle parameter $\vartheta_k=0$\cite{Rewoldt1982}, which
corresponds to the solution localized at the outside mid-plane,
i.e., $\theta_p=0$ in our notation, where $\theta_p$ is defined as
the local peaking poloidal angle for the mode structure.
For this reason, many local eigenvalue codes such as
HD7\cite{Dong2004} assume implicitly $\vartheta_k=0$. The
unconventional eigen modes with $\theta_p\neq0$ have been
recently discovered in the strong gradient parameter regime.
Typically, $|\theta_p|\simeq$ or $<$ $\pi/2$ have been shown to
exist\cite{Xie2012,Dickinson2014,Singh2014,Fulton2014}. In this
work, we find the most general unconventional eigen mode structures
from first principle gyrokinetic simulations. The underlying physics
is also explained and it has important implications for turbulent
transport.

\begin{figure}
\centering
\includegraphics[width=8.5cm]{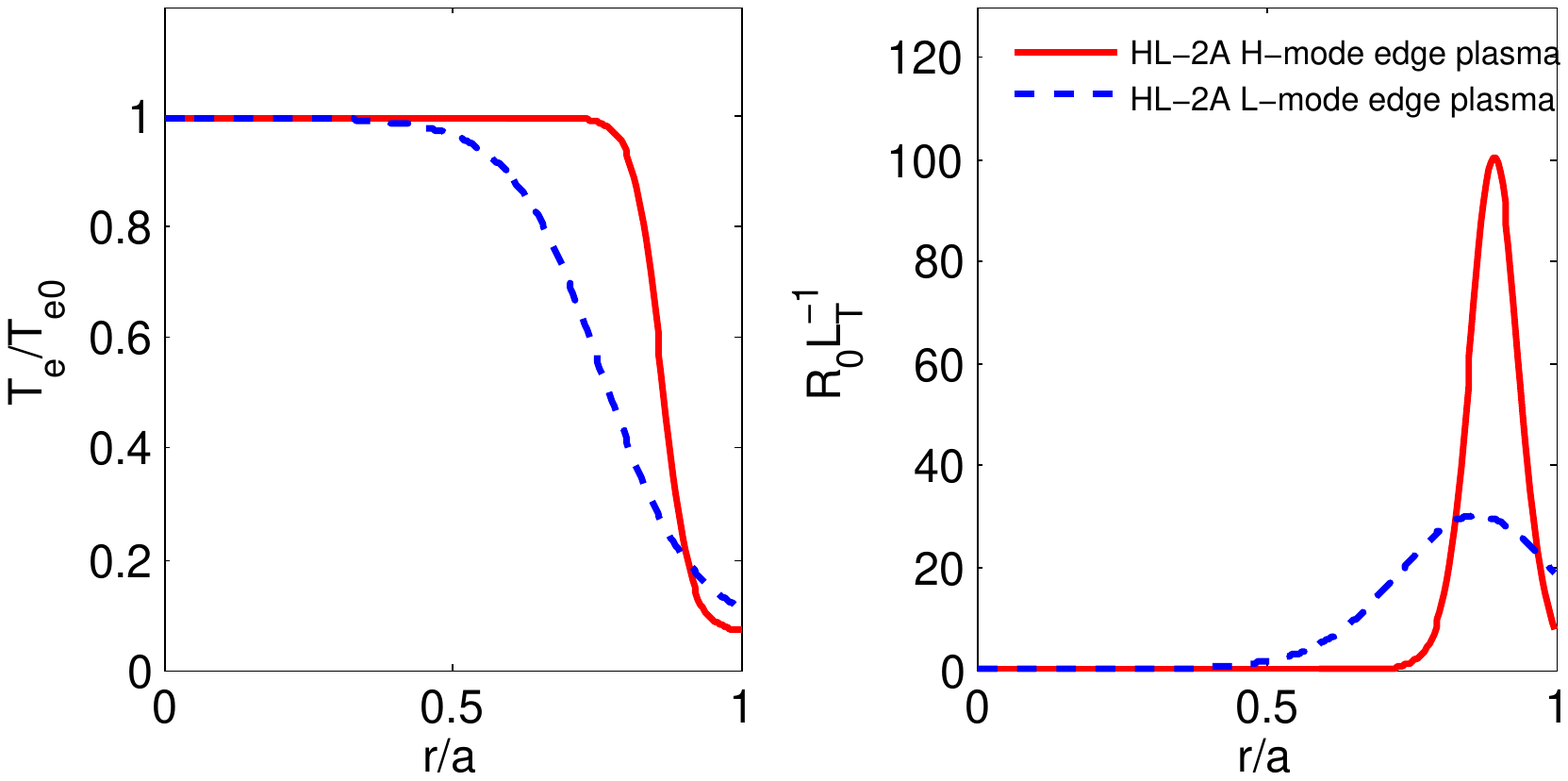}\\
\caption{Typical electron temperature (also density)
profiles used in this Letter. HL-2A L-mode edge plasma profile is
weak gradient $R_0L_T^{-1}<40$. HL-2A H-mode edge plasma profile is
strong gradient $R_0L_T^{-1}>80$.}\label{fig:profile_LH}
\end{figure}

We first obtain linear electrostatic results from global gyrokinetic
particle simulation using the GTC code\cite{Lin2004, Xiao2015} with
single toroidal mode number $n$. The simulation parameters and
profiles are similar to that of the recent H-mode experiments of the
HL-2A tokamak\cite{Xie2015b,Xie2015c}: toroidal magnetic field
$B_0=1.35T$, minor radius $a=40cm$, major radius $R_0=165cm$, safety
factor $q=2.5-3.0$, magnetic shear $s=0.3-1.0$, $R_0/L_n=80-160$
with $T_e(r)=T_i(r)$, and $n_e(r)=n_i(r)$. $L_n\equiv -(1/n)(dn/dr)$
and $L_T\equiv -(1/T)(dT/dr)$ are density and temperature gradient
scale length. Typical electron temperature (also density) profiles
used in this Letter are shown in Fig.\ref{fig:profile_LH}. We start
with $\eta=L_n/L_T=1.0$ for simplicity. Collisions are included in
some cases but shown little influence to the general results. Under
these parameters, no instability or only weakly unstable mode can be
found when the electrons are adiabatical. Thus, the major
instability for these simulation parameters is the trapped electron
mode.

\begin{figure}
\centering
\includegraphics[width=8.5cm]{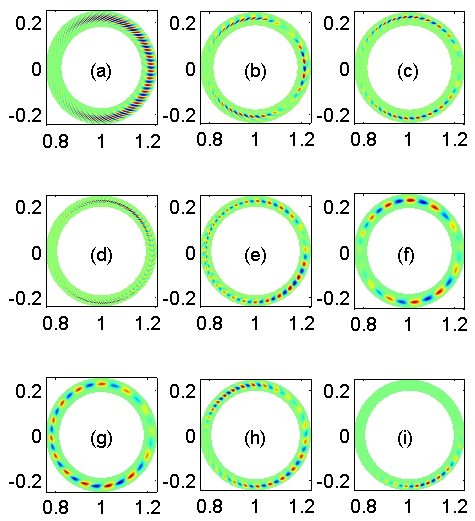}\\
\caption{Conventional (a) and unconventional (b-i) 2D ballooning
structures of electrostatic potential in (X,Z) plane for TEM
observed in GTC simulation, where (a) uses HL-2A tokamak edge weak
gradient L-mode plasma parameter ($R_0L_{n}^{-1}<40$) and (b)-(i)
use edge strong gradient H-mode parameters ($R_0L_{n}^{-1}>80$).
Collisions are only included in (e) and (g).}\label{fig:gtc_tem}
\end{figure}

\begin{figure}
\centering
\includegraphics[width=8.5cm]{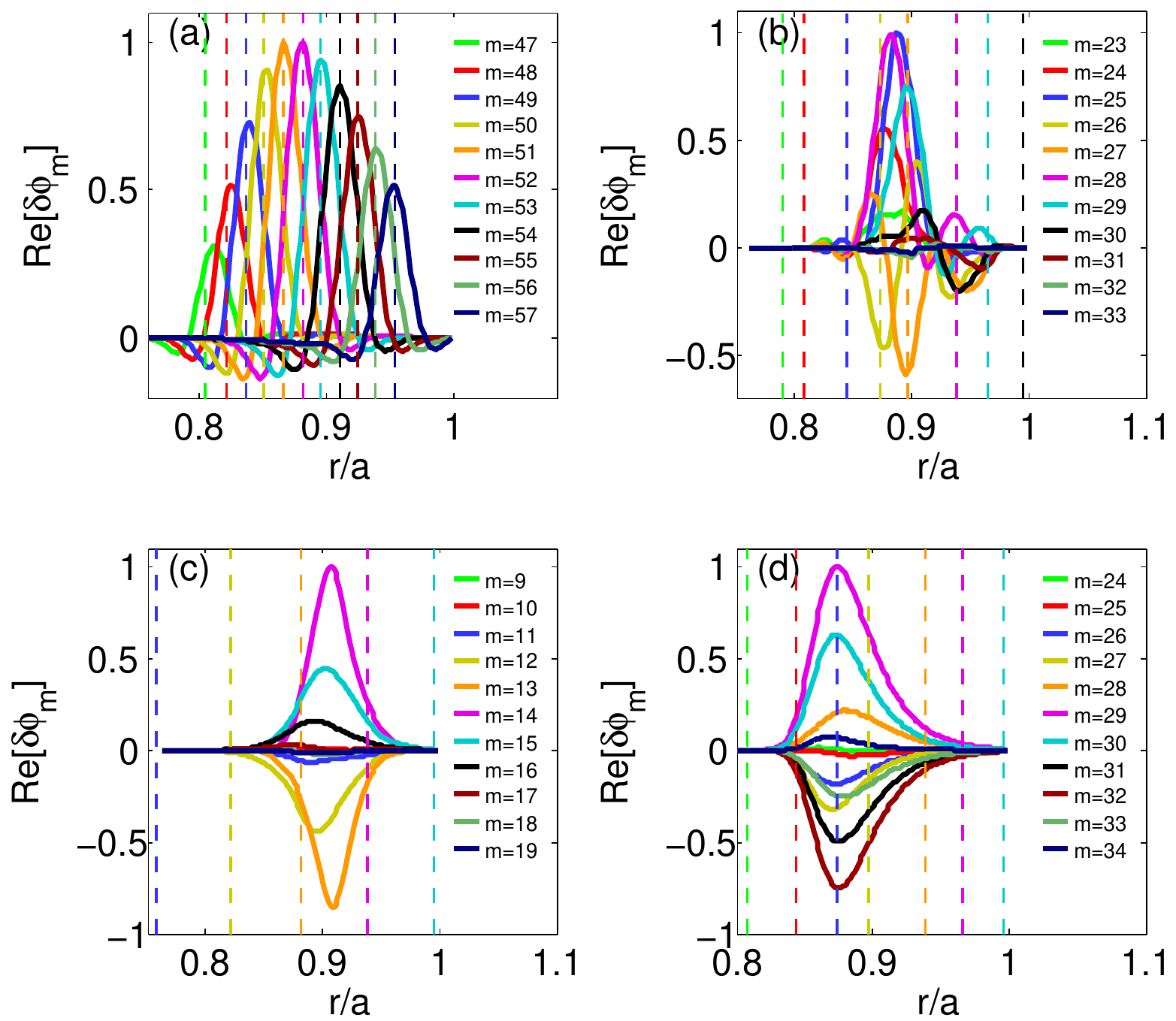}\\
\caption{The real part of Fourier $\delta\phi_m(r)$ for conventional
and unconventional mode structures. The corresponding poloidal cross
section mode structures of (a)-(d) ($n=20,10,5,10$,
respectively)  are taken from Fig.\ref{fig:gtc_tem} (a), (b), (g)
and (i), respectively. The dash lines are corresponding
rational surface positions $r_s$, where
$nq(r_s)=m$.}\label{fig:gtc_tem_phim}
\end{figure}

These TEM simulations show that both conventional and unconventional
ballooning mode structures can exist for various gradients and
toroidal mode numbers ($n=5-30$) , as shown by
Fig.\ref{fig:gtc_tem}. In these sub-figures, $q$ profiles are
similar. For Fig.\ref{fig:gtc_tem}b-i, the global
density (also temperature) profiles and toroidal mode
numbers are not the same but all are under strong gradient. The
novel features include: a). the mode can have anti-ballooning
structure (i.e., $|\theta_p|>\pi/2$, e.g., Fig.\ref{fig:gtc_tem}g);
b). the mode can have multiple peaks (e.g., Fig.\ref{fig:gtc_tem}b).
Considering that the trapped particles are mainly located at the low
magnetic field side, i.e., the outboard side, the anti-ballooning
structures of TEM are not expected. The 3D mode structure of the
electrostatic potential can be represented by the Fourier series
$\delta\phi(r,\theta,\zeta)=e^{in\zeta}\sum_m
\delta\phi_m(r)e^{-im\theta}$, where $m$ is poloidal mode number. To
explore the formation of these different eigenmode structures, we
compute the $\delta\phi_m(r)$ for several typical conventional and
unconventional mode structures, as shown in
Fig.\ref{fig:gtc_tem_phim}. For the conventional ballooning
structure, the poloidal eigen modes $\delta\phi_m(r)$ are almost
radially symmetric (Gaussian-like) and positive in amplitude. And,
$\delta\phi_m$ has a large overlap with $\delta\phi_{m+1}$, i.e.,
$\delta\phi_m\simeq\delta\phi_{m+1}$. However, for the
unconventional structures, the poloidal eigen modes
$\delta\phi_m(r)$ can be radially either symmetric or asymmetric,
and the amplitude for each symmetric mode can be either positive or
negative, as shown by Fig. 2b, c and d. Under stronger gradients the
radial peaking position of $\delta\phi_m(r)$ is also not at the
corresponding rational surface position $r_s$ any more, where
$nq(r_s)=m$.

\begin{figure}
\centering
\includegraphics[width=8.5cm]{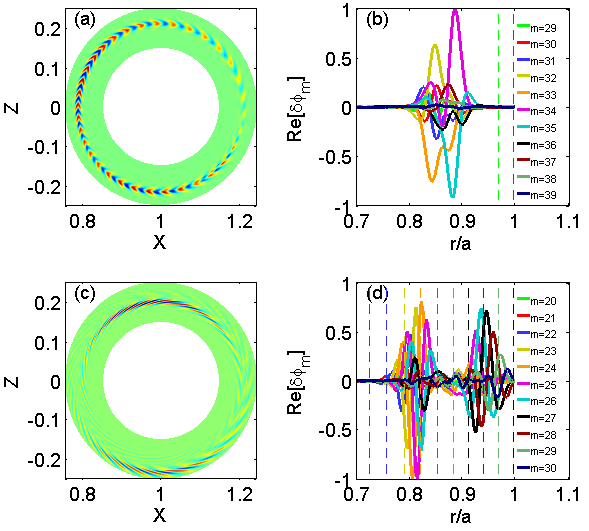}\\
\caption{Unconventional ITG ($n=10$)  mode structures in GTC. (a \&
b) Anti-ballooning structure. (c \& d) Two modes co-exist (or, one
mode with two radius peaks) at different radius positions. One has
$\theta_p\simeq\pi/2$ and another has
$\theta_p\simeq-\pi/2$.}\label{fig:gtc_itg}
\end{figure}

Next we consider ITG mode by reducing the density
gradient to $R_0/L_{n}<40$ and keeping the other parameters the same
as those for the TEM case, e.g., $R_0/L_{T}>80$ and thus
$\eta_i=L_n/L_T>2.0$. To completely exclude the contribution of
the kinetic electrons, we use adiabatic electrons in the
simulations. It is  found that the preceding unconventional mode
structures still exist and exhibit even more structural variations.
For example, the anti-ballooning structure is found for this ITG
simulation, as is shown in Fig.\ref{fig:gtc_itg}a\&b. Actually, the
mode structure with global profiles and multi modes coexisting in
the initial value simulation can be even more complicated. For
example, two modes with similar growth rates can be excited in
different radial locations, as shown in Fig.\ref{fig:gtc_itg}c\&d.
Multi modes coexist with close peaking positions in the initial
value simulation can also lead to $\theta_p=\theta_p(t)$, i.e.,
rotate poloidally with time. Thus, the unconventional mode
structures are not limited to TEM and can be common for drift waves.

These unconventional linear behaviors can be understood from the
following eigenmode analysis. We start with the ITG eigen mode
equation\cite{Connor1987,Dickinson2014}
\begin{eqnarray}\label{eq:itg2d}
&\Big[\rho_i^2\frac{\partial^2}{\partial
x^2}-\frac{\sigma^2}{\omega^2}\Big(\frac{\partial}{\partial
\theta}+ik_\theta
sx\Big)^2-\frac{2\epsilon_n}{\omega}\Big(\cos\theta+\frac{i\sin\theta}{k_\theta}\frac{\partial}{\partial
x}\Big)\nonumber\\
&-\frac{\omega-1}{\omega+\eta_s}-k_\theta^2\rho_i^2\Big]\delta\phi(x,\theta)=0,
\end{eqnarray}
where $\sigma=\epsilon_n/(qk_\theta\rho_i)$, $\epsilon_n=L_n/R_0$,
$\eta_s=1+\eta_i$, $x=r-r_s$, $r_s$ is the rational
surface, $\omega=\omega_r+i\gamma$ is the complex mode frequency
normalized by the electron diamagnetic frequency, and the poloidal
wave number $k_\theta=nq/r$. Eq.(\ref{eq:itg2d}) can be derived from
the gyrokinetic theory with adiabatic electron assumption. The
corresponding 1D eigen mode equation in the ballooning space is
\begin{eqnarray}\label{eq:itg1d}
&\Big\{\frac{\sigma^2}{\omega^2}\frac{d^2}{d
\vartheta^2}+k_\theta^2\rho_i^2[1+s^2(\vartheta-\vartheta_k)^2]+\frac{2\epsilon_n}{\omega}[\cos\vartheta\nonumber\\
&+s(\vartheta-\vartheta_k)\sin\vartheta]+\frac{\omega-1}{\omega+\eta_s}\Big\}\delta\hat\phi(\vartheta,\vartheta_k)=0,
\end{eqnarray}
where $\vartheta_k$ is the ballooning-angle parameter, which
represents an as yet undetermined radial
wavenumber\cite{Connor1987}. The relation between the
ballooning space electrostatic potential
$\delta\hat\phi(\vartheta,\vartheta_k)$ and real space
$\delta\phi(x,\theta)$ can be found in
Ref.\cite{Connor1987,Dickinson2014}. Using the Fourier basis
$\delta\phi(x,\theta)=\sum_m \delta\phi_me^{-im\theta}$,
Eq.(\ref{eq:itg2d}) can be rewritten as the 2D eigenmode equation
\begin{eqnarray}\label{eq:itg2dm}
&k_\theta^2\rho_i^2s^2\frac{\partial^2\delta\phi_m}{\partial
z^2}+\frac{\sigma^2}{\omega^2}(z-m)^2\delta\phi_m-\frac{\epsilon_n}{\omega}\Big[\Big(1-s\frac{\partial}{\partial
z}\Big)\delta\phi_{m-1}\nonumber\\
&+\Big(1+s\frac{\partial}{\partial
z}\Big)\delta\phi_{m+1}\Big]-\Big(\frac{\omega-1}{\omega+\eta_s}+k_\theta^2\rho_i^2\Big)\delta\phi_m=0,
\end{eqnarray}
where $z=k_\theta sx$. To solve the eigenvalue problem of
Eq.(\ref{eq:itg2dm}), only a few number of $m$ modes need
to be kept for the solution to reach convergence.

With suitable approximations (cf. Ref.\cite{Horton1981}),
both Eqs.(\ref{eq:itg1d}) and (\ref{eq:itg2dm}) can be reduced to
the Weber equation $u''+(bx^2+a)u=0$ (here the argument
$x$ is $\vartheta$ and $z$  for Eqs.(\ref{eq:itg1d}) and
(\ref{eq:itg2dm}) respectively), which has solutions with the
eigenvalues $a(\omega)=i(2l+1)\sqrt{b(\omega)}$ and eigenfunctions
$u(x)=H_l(i\sqrt{b}x)e^{-ibx^2/2}$, where $H_l$ is $l$-th Hermite
polynomial and $l=0,1,2,...$, which represent a series eigenstates.
With the original equations, i.e., Eqs.(\ref{eq:itg1d}) and
(\ref{eq:itg2dm}), which can only be solved numerically, the
eigenstates take a more complicated form.

Eqs.(\ref{eq:itg1d}) and (\ref{eq:itg2dm}) can be solved numerically
by transforming it to a matrix eigenvalue problem as $\omega^3{\bm
M_3}{\bm X}+\omega^2{\bm M_2}{\bm X}+\omega{\bm M_1}{\bm X}+{\bm
M_0}{\bm X}=0$, with X is the discrete representation of
the electrostatic potential. We use finite difference to discretize
the system, which yields sparse matrices for ${\bm M_i}$
($i=0,1,2,3$). Using the companion matrix method, the nonlinear
eigenvalue problem can be transformed to a standard eigenvalue
problem as ${\bm A}{\bm Y}=\omega{\bm B}{\bm Y}$, where ${\bm
Y}=[{\bm X_1},{\bm X_2},{\bm X_3}]\equiv[{\bm X},\omega{\bm
X},\omega^2{\bm X}]$, ${\bm A}=[{\bm O},{\bm I},{\bm O};{\bm O},{\bm
O},{\bm I};-{\bm M_0},-{\bm M_1},-{\bm M_2}]$, ${\bm B}=[{\bm
I},{\bm O},{\bm O};{\bm O},{\bm I},{\bm O};{\bm O},{\bm O},{\bm
M_3}]$, and ${\bm I}$ and ${\bm O}$ are unit and null matrix
respectively. Thus all the solutions of this eigen system can be
obtained (c.f., \cite{Xie2015} for details of similar treatment).
The advantage of this method is that it can show the complete set
solutions of the discrete eigen system and help us to understand the
distribution of eigenvalues in the complex plane. The solution in
Refs.\cite{Xie2012,Dickinson2014} using iterative solver is actually
just one of the solutions obtained here and may not be the most
unstable or most important, which depends heavily on the initial
guess. This companion matrix method has been verified by comparing
the numerical solutions with that from the shooting method and the
analytical solution for the Weber equation.

\begin{figure}
\centering
\includegraphics[width=8.5cm]{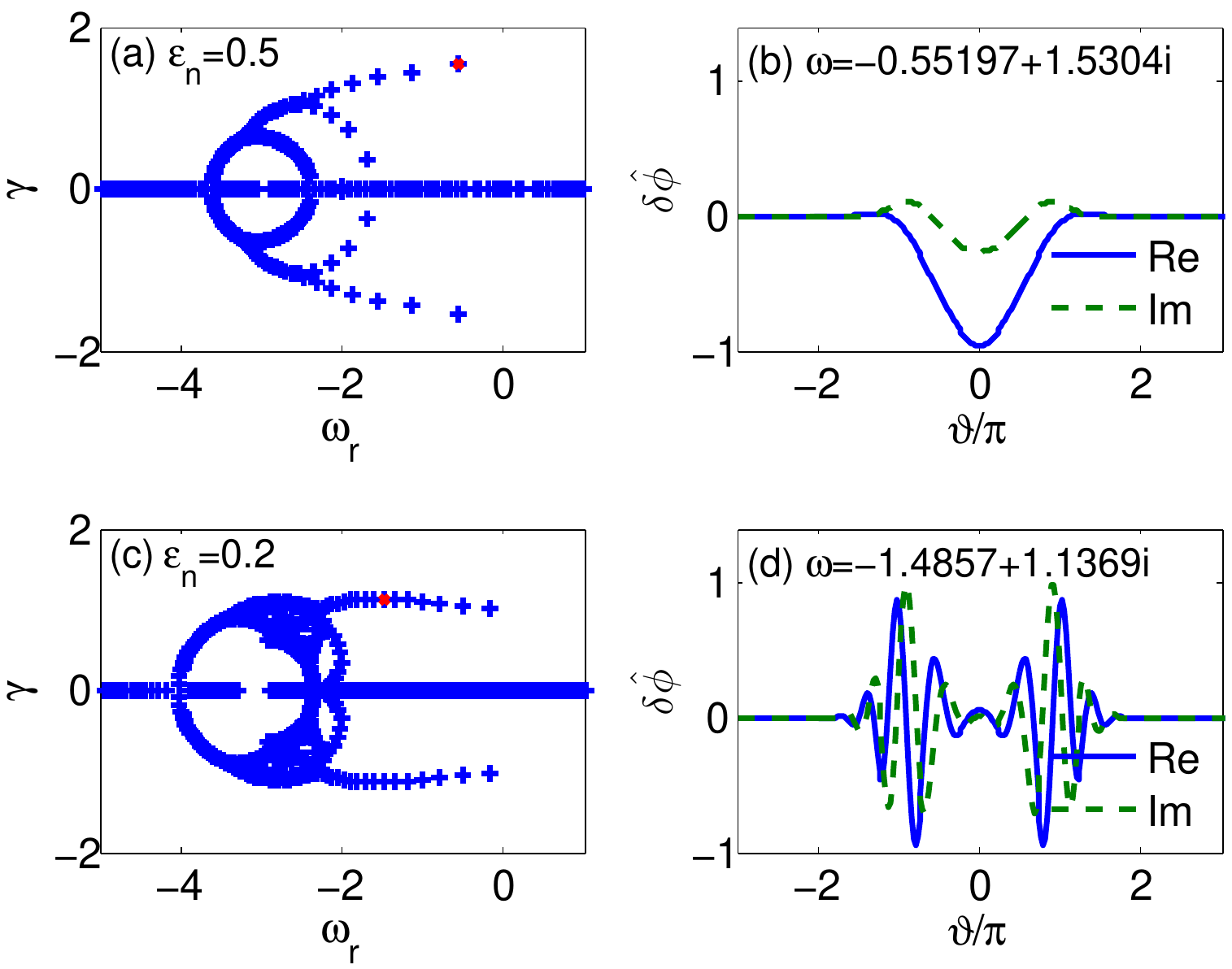}\\
\caption{In Eq.(\ref{eq:itg1d}), series of solutions
exist. For weak gradient ($\epsilon_n=0.5$), the most unstable
solution (red 'x') is the ground state (a\&b), which is the
conventional ballooning structure. For strong gradient
($\epsilon_n=0.2$), the most unstable solution (red 'x') is not the
ground state (c\&d), which represents the unconventional ballooning
structure.}\label{fig:itg1d}
\end{figure}

By solving Eq.(\ref{eq:itg1d}) in the 1D ballooning space, the
unconventional ballooning structures occur when either the most
unstable solution is not the ground eigen state ($l\neq0$), or the
ballooning angle $\vartheta_k\neq0$. Both of these two conditions
can be met in the strong gradient regime. The most unstable solution
with $\vartheta_k\neq0$ has been discussed by others (c.f.,
\cite{Singh2014,Lu2015}). Here we focus on the unconventional
ballooning structure caused by the non-ground eigen state. The
following parameters are used to solve Eq.(\ref{eq:itg1d}): $s=0.8$,
$k_\theta\rho_i=0.4$, $q=1.0$, $\eta_s=3.0$ and $\vartheta_k=0$. As
is known from the aforementioned analytical analogy,
Fig.\ref{fig:itg1d} shows that a series of solutions can exist for
Eq.(\ref{eq:itg1d}), where $R/L_T$ and $R/L_n$ are changed
simultaneously to ensure $\eta=L_n/L_T=const.$, and one should also
note that the frequency is normalized by the electron diamagnetic
frequency. For the weak gradient case ($\epsilon_n=0.5$), we find
that the most unstable solution is the ground state
(Fig.\ref{fig:itg1d}a), which is the conventional ballooning
structure (Fig.\ref{fig:itg1d}b). For the strong gradient case
($\epsilon_n=0.2$), the most unstable solution is not the ground
state (Fig.\ref{fig:itg1d}c\&d), which corresponds to the
unconventional ballooning structure. More detailed
analysis\cite{Chen2015} of Eq.(\ref{eq:itg1d}) for present
discussion of the unconventional mode structure can be obtained by
extension of Refs.\cite{Chen1980,Horton1981}.

\begin{figure}
\centering
\includegraphics[width=8.5cm]{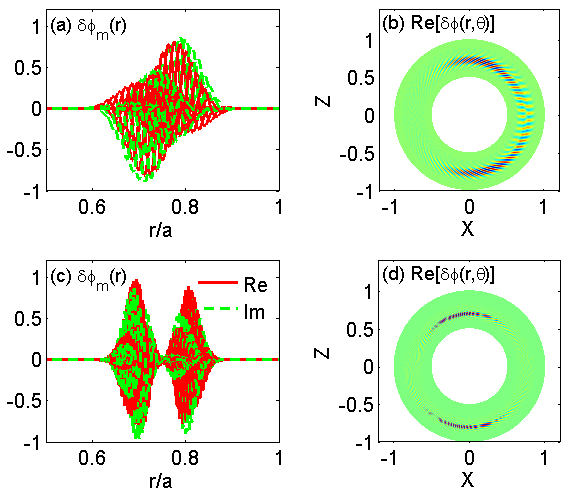}\\
\caption{Typical unconventional mode structures from 2D eigen
solution for Eq.(\ref{eq:itg2dm}). (b) is similar to
Fig.\ref{fig:gtc_tem}(c\&d), and (c\&d) are similar
to Fig.\ref{fig:gtc_itg}(d\&c)}\label{fig:itg2d}
\end{figure}

We have demonstrated that, with strong gradient the most unstable
solution can shift from ground state to other non-ground states,
which is analogous to the quantum jump between energy levels.
Physically, the jump behavior can be understood from the effective
potential\cite{Chen1980}. The jump happens from one potential well
to another, which leads to different energy levels. It is not
transparent that the non-ground eigen state in the 1D ballooning
space corresponds to the unconventional mode structure in the 2D
poloidal plane. Next we confirm this link by showing that the
non-ground 2D eigen state solved from Eq.(\ref{eq:itg2dm}) can form
the unconventional mode structures observed in the preceding
gyrokinetic simulation. The solutions in
Refs.\cite{Xie2012,Dickinson2014,Singh2014} are just a weak
asymmetric solutions of our series solutions. Almost all the mode
structures in Figs.\ref{fig:gtc_tem} and \ref{fig:gtc_itg} have also
been found in the 2D eigen solutions of Eq.(\ref{eq:itg2dm}). Two
examples are shown in Fig.\ref{fig:itg2d}. Therefore, conventional
and unconventional series solutions have been found in both 2D eigen
solver and GTC initial simulations. The condition for the jump of
the most unstable eigen state to non-ground state is
$\epsilon_n<\epsilon_c$, where $\epsilon_c$ is a critical gradient
parameter which depends on other parameters. In GTC simulations of
the HL-2A parameters, the typical critical density (or
temperature) gradient value is $R_0/L_n=40-120$.

The results from the gyrokinetic simulation and eigen mode analysis
show that the unconventional mode structures exist mainly in the
strong gradient regime or the H-mode. In the weak gradient regime or
L-mode, conventional mode structures still prevail. This can
indicate different transport behavior between H-mode and
L-mode\cite{Chen2014}. In the conventional ballooning structure, the
neighboring Fourier modes $\delta\phi_m\simeq \delta\phi_{m+1}$, the
effective correlation length may be estimated as the width of radial
envelope of the modes, say, $\Delta A$. Whereas, in the
unconventional ballooning structures, especially for anti-ballooning
structure, $\delta\phi_m\simeq -\delta\phi_{m+1}$ can occur, i.e., a
$180^{\circ}$ phase shift for the neighboring Fourier modes, which
can change the effective correlation length to the distance of
neighboring mode-rational surfaces $\Delta r_s$. Considering that
$\Delta r_s\ll\Delta A$, we can expect that the H-mode can have
better confinement.

To summarize, a broad class of unconventional ballooning modes are
found for electrostatic drift waves (TEM and ITG) by the gyrokinetic
simulation, which is shown to be common in the strong gradient
regime. These unconventional mode structures are shown to correspond
to the non-ground-state solutions of the eigen mode equation. These
results may have important implications for the turbulent transport
in tokamaks, i.e., the turbulent transport mechanism in the H-mode
can be rather different from that in the L-mode, which requires
further investigation by self-consistent nonlinear gyrokinetic
simulations\cite{Xie2015c}.

{\it Acknowledgments} Discussions with L. Chen, H. T. Chen, Z. X. Lu
and Z. Lin are acknowledged. The work was supported by the National
Magnetic Confinement Fusion Science Program under Grant Nos.
2015GB110000 and 2013GB111000, the Recruitment Program of Global
Youth Experts.

\end{document}